\documentclass[conference]{IEEEtran}
\IEEEoverridecommandlockouts

\usepackage{amsmath,amssymb}
\usepackage{geometry}
\usepackage{braket}
\usepackage{bbm}
\usepackage{hyperref}
\usepackage{physics}
\usepackage{glossaries}
\usepackage{xcolor}
\usepackage{graphicx}
\usepackage{cuted} 
\usepackage{wrapfig}

\newtheorem{remark}{Remark}

\newtheorem{definition}{Definition}
\newcommand{\eins}{\mathbbm{1}}

\newacronym{tmsv}{TMSV}{Two-Mode Squeezed Vacuum}
\newacronym{bc}{BC}{Bit Commitment}
\newacronym{ttp}{TTP}{Trusted Third Party}

\makeatletter
\renewcommand\paragraph{\@startsection{paragraph}{4}{\z@}%
	{1.25ex \@plus1ex \@minus.2ex}%
	{-.1em}%
	{\normalfont\normalsize\bfseries}}
\makeatother

\def\BibTeX{{\rm B\kern-.05em{\sc i\kern-.025em b}\kern-.08em T\kern-.1667em\lower.7ex\hbox{E}\kern-.125emX}}
\begin{document}
	
	\title{Phase-Based Bit Commitment Protocol
	}

	\author{
		\IEEEauthorblockN{Janis N\"otzel\IEEEauthorrefmark{1}, Anshul Singhal\IEEEauthorrefmark{1}, Peter van Loock\IEEEauthorrefmark{2}}
		\thanks{ Copyright © 2026 IEEE. This material may be used for personal purposes.
			Any other use, including reproduction or redistribution, requires prior
			permission from IEEE. Requests for such permission should be sent to
			pubs-permissions@ieee.org.\par
			This work has been accepted for presentation at IEEE ICC 2026.}
		\IEEEauthorblockA{
			\IEEEauthorrefmark{1}Emmy-Noether Group Theoretical Quantum Systems Design Lehrstuhl f\"ur Theoretische Informationstechnik,\\ Technische Universit\"at M\"unchen\\ 
			\{janis.noetzel, anshul.singhal\}@tum.de 
		}
		\IEEEauthorblockA{
			\IEEEauthorrefmark{2}Institut für Physik,\\Johannes Gutenberg Universit\"at Mainz    }}
	\maketitle

	\begin{abstract}
		With the rise of artificial intelligence and machine learning, a new wave of private information is being flushed into applications. This development raises privacy concerns, as private datasets can be stolen or abused for non-authorized purposes. Secure function computation aims to solve such problems by allowing a service provider to compute functions of datasets in the possession of a a data provider without reading the data itself. A foundational primitive for such tasks is \gls{bc}, which is known to be impossible to realize without added assumptions. Given the pressing nature of the topic, it is thus important to develop \gls{bc} systems and prove their security under reasonable assumptions. In this work, we provide a novel quantum optical \gls{bc} protocol that uses the added assumption that the network provider will secure transmission lines against eavesdropping. Under this added assumption, we prove security of our protocol in the honest but curious setting and discuss the hardness of Mayer's attack in the context of our protocol. 
	\end{abstract}

	\section{Introduction}

	\gls{bc} \cite{chailloux2011optimal} is a primitive task for cryptography having applications in various tasks such as oblivious transfer (OT) \cite{rabin2005exchange} and by extension two-party computation. The process involves two different parties, a sender and a receiver. The sender's task is to 'commit' to a single bit i.e., it can't change the chosen bit after sending to the receiver. This is also known as commit phase. However, it needs to provide some evidence to the receiver about the committed bit from which the receiver cannot identify the chosen bit. In other words, the bit is concealed. At a later stage, the sender has to reveal the chosen bit value, known as opening phase. In an honest protocol sender can't change the chosen bit value. 
	
	\indent Quantum mechanics offers new possibilities for cryptography, but early hopes for unconditional quantum \gls{bc} were dashed by seminal no-go theorems. Bennett and Brassard’s \cite{brassard1997brief} original 1984 protocol was shown to be insecure against entanglement-based cheating. In 1997 Mayers \cite{mayersImpossibility} and, independently, Lo and Chau \cite{lo1998quantum} proved that any quantum \gls{bc} protocol whose security is based solely on quantum laws is fundamentally insecure. Intuitively, if Alice initially prepares entangled EPR pairs instead of fixed states, she can delay her choice and later measure in one basis or the other to unveil either bit at will, with Bob unable to detect the difference. Thus, unconditionally secure quantum \gls{bc} (simultaneously concealing and binding) is impossible without further assumptions. To circumvent these no-go results the research focus shifted on to additional assumptions under which the secure \gls{bc} is possible. 
	
	\indent Assumption-based quantum protocols are both natural and necessary because realistic physical systems never realize the idealized, unrestricted adversary assumed in impossibility proofs. Practical devices have limitations such as finite coherence times, imperfect quantum memories, bounded entangling capabilities, limited communication bandwidth, and relativistic constraints on information transfer and these limitations can be framed as benign, physically motivated assumptions that restore cryptographic functionality. For example, one of the prominent restriction is by limiting the quantum memory \cite{konig2012unconditional,kaniewski2013secure}. In \cite{damgaard2008cryptography} Damgaard et al showed, honest parties need no quantum memory, whereas an adversary needs unusually large quantum storage to break the protocol. In \cite{wehner2008cryptography}, an approach based on the noisy storage model was investigated. If an attacker's quantum memory decoheres sufficiently between rounds, one can one achieve \gls{bc} with with unconditional security. In \cite{adlam2015deterministic} authors used relativistic constraints to achieve unconditional security. They developed protocols in which Alice and Bob coordinate between spatially separated laboratories such that any attempt of cheating would violate causality. They also described deterministic relativistic quantum \gls{bc} protocol relying on Minkowski causality and entanglement. Each of these methods adds assumptions that evade the standard no-go theorems. Adopting such assumptions shifts the security model from purely information-theoretic impossibility to a hybrid model that is both analyzable and experimentally relevant: it permits rigorous security proofs under explicit physical constraints and suggests concrete implementation requirements (memory size, coherence time, synchronization accuracy) that experimentalists can target. Thus, assumption-based protocols offer a pragmatic pathway to realizing advanced two-party primitives (like \gls{bc} and oblivious transfer) in the quantum era, providing a bridge between theoretical limits and physically implementable security. The works of \cite{Chiribella_2013,approximateBC} expanded the scope of the known no-go results to further protocol classes and more generic quantum systems. In \cite{chaoui2025securequantumbitcommitment}, the authors showed that quantum \gls{bc} becomes possible when only separable operations are allowed for the committing party.
	
	In this work we provide a novel protocol starting out from the assumption that network connections will be secured by the network provider. The provider thus becomes a \gls{ttp}. In a realistic use of our protocol where the transmittivity $\tau$ of the link connecting Alice and Bob satisfies $\tau<1$, the provider ensures that Bob cannot access the link at a different position and thereby change $\tau$ in his favor. In addition, our discussion will reveal that the provider could ensure $\tau\ll1$ for protocol efficiency. 
	\section{Notation and Definitions}
	\paragraph*{Basics}
	For every $M\in\mathbb N$ we set $[M]:=\{0,\ldots,M-1\}$ and let $\oplus$ denote addition module $M$ on $[M]$, and $\ominus$ the respective subtraction. When $\oplus$ or $\ominus$ are involved, $M$ will be implicit from the context. The set of probability measures on a finite set $\mathbf A$ is written $\mathcal P(\mathbf A)$, and the set of states on a Hilbert space $\mathcal H$ is $\mathcal P(\mathcal H)$. The trace of an operator $A$ on $\mathcal H$ is $\tr(A)$, the scalar product of $x,y\in\mathcal H$ $\langle x,y\rangle$. The Kronecker delta is $\delta_{mn}=1$ iff $m=n$ and $\delta_x=1$ iff $x=1$.\\
	The one-norm on $\mathcal P(\mathcal H)$ is denoted as $\|\cdot\|_1$. We note that it has an operational interpretation as the maximum probability of a given measurement for distinguishing two quantum states $\varrho,\varsigma$: $\|\varrho-\varsigma\|_1=2\max_{0\leq P\leq\eins}\|P(\varrho-\varsigma)\|_1$.\\
	For $a\in[0,1]$ we set $a':=1-a$. The single-mode Fock space is $\mathcal F$. It is spanned by the photon number basis $\{|n\rangle\}_{n=0}^\infty$ and has corresponding sub-spaces $\mathcal F_N:=\mathrm{span}\{|0\rangle,\ldots,|N\rangle\}$ with restricted photon numbers and corresponding projections $P_N:=\sum_{n=0}^{N}|n\rangle\langle n|$. On $\mathcal F$, we use the usual creation operators $\hat a^\dagger$ which act as $\hat a^\dagger|n\rangle = \sqrt{n+1}|n+1\rangle$.\\ 
	We will need another important type of state: The so-called \emph{coherent} states $|\alpha\rangle$. For a generic complex number $\alpha$, the corresponding coherent state is defined as $|\alpha\rangle:=\exp(-|\alpha|^2/2)\sum_{n=0}^\infty\tfrac{\alpha^{n}}{\sqrt{n!}}|n\rangle$. A displacement operator $D(\beta)$ ($\beta\in\mathbb C$) changes $|\alpha\rangle$ to $D(\beta)|\alpha\rangle = |\alpha+\beta\rangle$. A measurement of the photon number operator on $|\alpha\rangle$ yields result $n\in\mathbb N$ with probability $|\alpha|^{2n}/n!$.\\
	Finally, the  Wigner function of the state $\rho$ is a quasi-probability distribution on phase space $(x,p) \in \mathbbm{R}^{2n}$ which is defined as $W_{\rho}(x,p) = \frac{1}{(\pi \hbar)} \int_{-\infty}^{\infty} \langle x-y | \hat{\rho} | x+y \rangle  e^{2ipy/\hbar} dy$. The Wigner function of a convex combination $\int p(\alpha)|\alpha\rangle\langle\alpha|d\alpha$ is always non-negative and equal to the convolution of $p$ with a Gaussian: $W(\rho)(x,p)=\int p(\alpha)e^{-|\alpha-x-\mathbbm{i}p|^2}d\alpha$. In this case, they describe the statistics of a homodyne measurement and bear operational significance.
	
	\begin{definition}[$\epsilon$-Security]
		A bit commitment protocol needs to be \emph{binding} (implying that Alice cannot change her mind after she committed to a bit $b$) and \emph{concealing} (meaning that Bob cannot learn the value of $b$ before Alice reveals it). The protocol that we propose here is only approximately binding and concealing. 
		In order to assess its quality via a single number we define the notion of $\epsilon $-security. The protocol is said to be $\epsilon$ secure if the probability that Bob can learn $b$ before Alice reveals it is upper bounded by $\epsilon$ and at the same time the probability that Alice can cheat and change her mind regarding her commitment is upper bounded by $\epsilon$. 
	\end{definition}
	\begin{remark}[Trusted Third Party]
		For sake of simplicity, we assume $\tau=1$ in our technical derivations. 
	\end{remark}
	
	\section{Protocol}
\begin{figure}
		\centering
		\begin{minipage}[t]{.24\textwidth}
			\includegraphics[width=\textwidth]{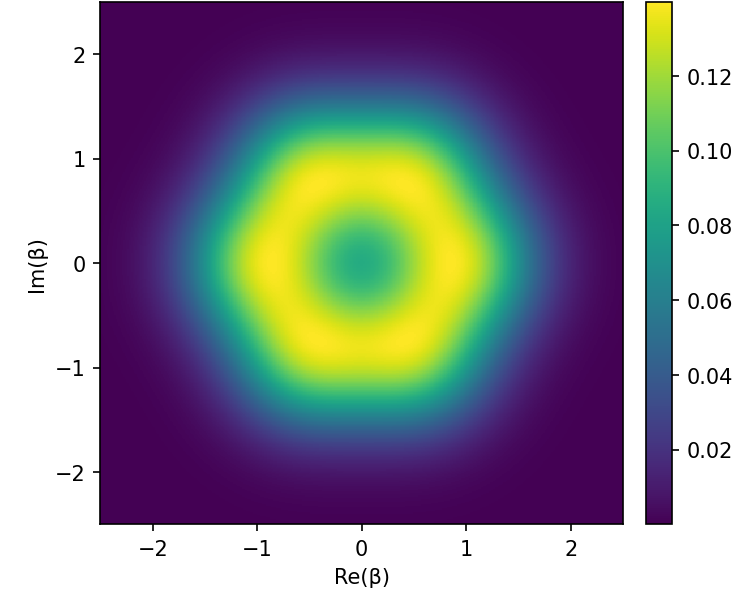}
		\end{minipage}\begin{minipage}[t]{.24\textwidth}
			\includegraphics[width=\textwidth]{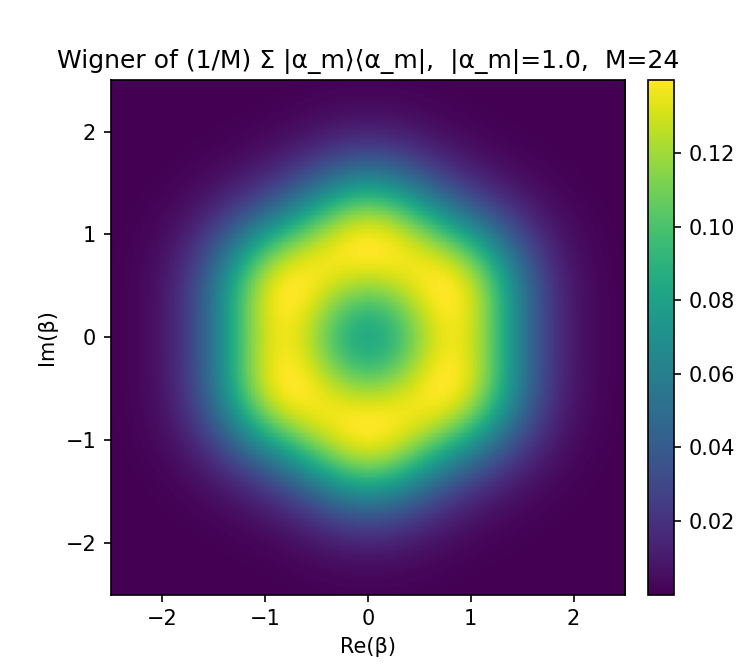}
		\end{minipage}
		\begin{minipage}[t]{.24\textwidth}
			\includegraphics[width=\textwidth]{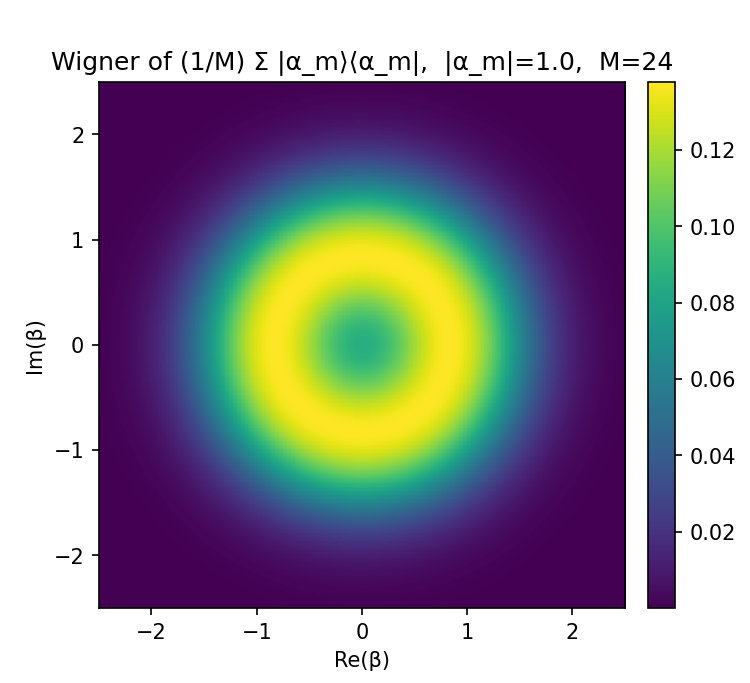}
		\end{minipage}\begin{minipage}[t]{.24\textwidth}
			\includegraphics[width=\textwidth]{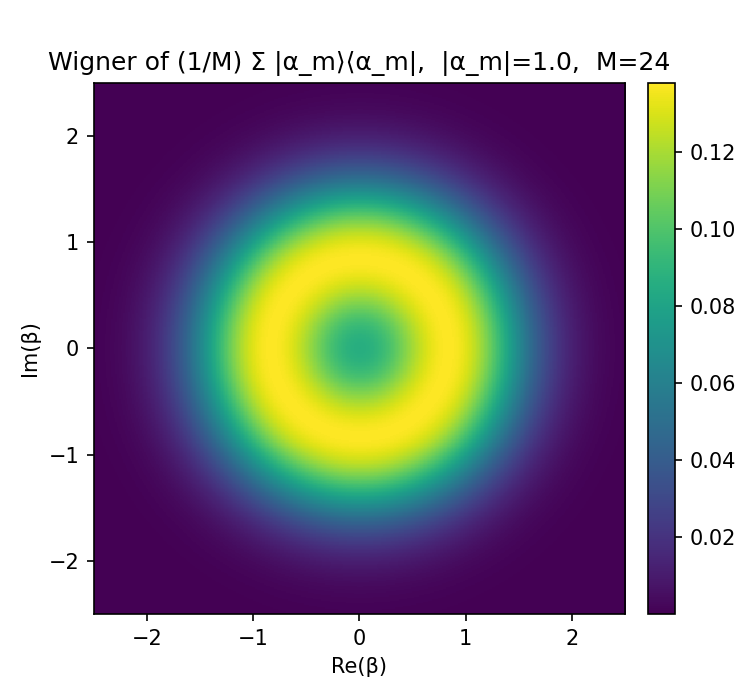}
		\end{minipage}
		\caption{Exemplary Wigner plots of the code states at Bob's end. Examples are made by setting $t=1$. The top row shows the choice $b=0$ (left) and $b=1$ (right) for $M=6$. The bottom row shows the same choices of $b$ for $M=32$. Intuitively, both choices can be easily differentiated for $M=6$, but are hardly distinguishable for $M=32$.}\label{fig:visualization}
	\end{figure}

	The protocol assumes system parameters such as the \emph{received} energy $E$ of the signals and the number $k$ of signals which together form a commitment are fixed and agreed upon between Alice and Bob.
	\paragraph*{Commitment}
	Alice commits to a bit $b$ and samples a random string $m^k\in[M]^k$. She then prepares the state
	\begin{align}
		|\psi_b(m^k)\rangle:=\otimes_{j=1}^k|\sqrt{E}e^{\mathbbm{i}2\pi\cdot (m_j+\tfrac{1}{2}b)/M}\rangle
	\end{align}
	and sends this state to Bob. 
	\paragraph*{Opening}
	Alice sends $(b,m^k)$ to Bob. He displaces the state he received from Alice by applying displacements $D(-2\pi\cdot (m_1+\tfrac{1}{2}b)/M)\otimes\ldots\otimes D(- 2\pi\cdot (m_k+\tfrac{1}{2}b)/M)$. He then measures in the photon number basis. Upon measuring the result $A=(0,\ldots,0)$ he accepts Alice's claim. If his measurement result is not equal to $A$ he rejects and the protocol aborts with an error.
	\begin{remark}
		After receiving Alice's commitment and without any extra knowledge, Bob holds either one of $\sigma_0^{\otimes k}$ or $\sigma_1^{\otimes k}$. Upon choosing the appropriate parameters $E,k$ and $M$ the protocol ensures that both code states $\sigma_0^{\otimes k},\sigma_b^{\otimes k}$ are almost identical to $\rho^{\otimes k}$. This ensures that the protocol is hiding. To understand why it is also binding, note that after Alice reveals $(b,m^k)$ Bob's state changes to a specific coherent state. Assuming Bob is equipped with a perfect photon counting device plus the ability to perform a displacement operation, he can test whether the state he holds equals the one that he should have if Alice did not cheat. This measurement will succeed with high probability, making it hard for Alice to cheat. 
		We visualize the functionality of the protocol in Figure \ref{fig:visualization}.
	\end{remark}
	
	\paragraph*{Ideal Phase Averaged State} 
	In our proposed protocol, a key ingredient will be to show that Alice's commitments are both almost indistinguishable from one and the same quantum state. This state is the so-called ``ideal averaged state'' that will, at Bob's side, be represented by a phase-averaged coherent state:
	\begin{align}
		\rho = \frac{1}{2\pi}\int_0^{2\pi}\ket{e^{2\pi\mathbbm{i}\theta}\alpha}\!\bra{e^{2\pi\mathbbm{i}\theta}\alpha}\,d\theta.
	\end{align}
	Using $\int_0^{2\pi} e^{i(m-n)\theta}\,d\theta = 2\pi\,\delta_{mn}$, we get for $|\alpha|^2=E$
	\begin{align}
		\bra{m}\rho\ket{n}
		= e^{-E} \frac{\sqrt{E}^{m+n}}{\sqrt{m!\,n!}} \,\delta_{mn}
		= e^{-E}\frac{\sqrt{E}^{2n}}{n!}\,\delta_{mn}.
	\end{align}
	Therefore, $\rho$ is diagonal in the number state basis.
	\noindent\paragraph*{Average Code State}
	\begin{align}\label{eqn:average-code-state}
		\sigma_b = \frac{1}{M}\sum_{k=0}^{M-1}\ket{\sqrt{E}e^{2\pi\mathbbm{i}\tfrac{k+b/2}{M}}}\!\bra{\sqrt{E}e^{2\pi\mathbbm{i}\tfrac{k+b/2}{M}}}
	\end{align}
	The corresponding matrix elements are then as follows:
	\begin{align}
		\bra{m}\sigma_b\ket{n}
		&= \frac{e^{-E}\sqrt{E}^{m+n}}{M\sqrt{m!\,n!}}
		\sum_{k=0}^{M-1} e^{2\pi i (k+\tfrac{b}{2})\tfrac{m-n}{M}}\\
		&= e^{-E}\frac{\sqrt{E}^{m+n}}{\sqrt{m!\,n!}}e^{\mathbbm{i}\pi b(m-n)/M}
		\,\delta_{m\ominus n}.
	\end{align}
	The Wigner function of $\sigma_0$ and $\sigma_1$ in Figure \ref{fig:visualization} for different values of $M$. 
	
	\section{Security}
	We first study the protocol in the case where Bob may perform any measurement to learn Alice's commitment before the reveal stage, but Alice has to follow all steps of the protocol and is only allowed to deviate with respect to the values that she reveals. We then investigate the relation to Mayers attack. In order to streamline notation, we will use in what follows the abbreviation $t:=\sqrt{E}$.
	\paragraph*{Cheating Bob} 
	Before the opening, Bob does not know $(b,m^k)$. He therefore holds $k$ copies of $\rho_b$ as defined in \eqref{eqn:average-code-state}.
	
	Let without loss of generality $b=0$ and set $D = \rho - \sigma_0$. Its matrix elements are
	\begin{align}
		D_{mn}
		= e^{-t^2}\frac{t^{m+n}}{\sqrt{m!\,n!}}e^{\mathbbm{i}\pi \tfrac{b(m-n)}{M}}
		\bigl[\delta_{mn} - \delta_{m\ominus n=0}\bigr].
	\end{align}
	We observe that the terms $\delta_{m\ominus n=0}$ only give a contribution when $m=n+k\cdot M$ for some integer value $k$. We use this fact to give a simple upper bound on the trace norm of the difference operator $D$:
	\begin{align}
		\|D\|_1&\leq e^{-t^2}\sum_{m,n=0}^\infty|D_{m,n}| = e^{-t^2}\sum_{\substack{m> n}}\delta_{n\ominus m}\frac{t^{n+m}}{\sqrt{n!m!}}\nonumber\\
		&=e^{-t^2}\sum_{n=0}^\infty\sum_{k=0}^\infty\frac{t^{n+n+kM}}{\sqrt{n!}\sqrt{(n+kM)!}}.
	\end{align}
	We proceed by utilizing log convexity of the factorial \cite{logConvexity}, which yields
	\begin{align}
		\|D\|_1\leq e^{-t^2}\sum_{n=0}^\infty\sum_{k=1}^\infty\frac{t^{n+n+kM}}{(n+kM/2)!}.
	\end{align}
	Application of the inequality $(n+m)!\geq n!m!$ yields
	\begin{align}
		\| D\|_1 \leq e^{-t^2}\sum_{n=0}^\infty\sum_{k=1}^\infty\frac{t^{n+n+kM}}{n!(kM/2)!}.
	\end{align}
	By applying Stirling's approximation in the form $n!\geq n^ne^{-n}\sqrt{2\pi n}e^{\tfrac{1}{12n+1}}\geq n^ne^{-n}$ and the formula for the geometric series, we get
	\begin{align}
		\|D\|_1&\leq e^{-t^2}\sum_{n=0}^\infty\sum_{k=1}^\infty\frac{t^{n+n+kM}}{n!(kM/2)!}\\
		&\leq e^{-t^2}\sum_{n=0}^\infty\frac{(t^{2})^n}{n!}\sum_{k=1}^\infty\frac{(t^2)^{kM/2}}{(kM/2)!}.
	\end{align}
	Finally, this gives
	\begin{align}
		\|D\|_1&\leq\sum_{k=1}^\infty\frac{(t^2)^{kM/2}}{(kM/2)!}\leq 2\left(\tfrac{2et^2}  {M}\right)^{M/2}.
	\end{align}   
	Here, the last second inequality is only valid if $(2et^2/M)^{M/2}<1/2$. If even $M>4et^2+1$, we get 
	\begin{align} 
		\|D\|_1 \leq 2^{-M/2}.
	\end{align}
	Bob's cheating probability $p_{CB}$ is equal to his probability of distinguishing between $b=0$ and $b=1$ before the opening. Using the operational interpretation of the one norm \cite{nielsen-chuang} and the telescope sum identity, it is upper bounded as
	\begin{align} 
		p_{CB}&\leq\tfrac{1}{2}\|\sigma_0^{\otimes k}-\sigma_1^{\otimes k}\|_1\\
		&\leq\tfrac{1}{2}\|\sigma_0^{\otimes k}-\rho^{\otimes k}\|+\tfrac{1}{2}\|\rho^{\otimes k}-\sigma_1^{\otimes k}\|_1\\
		&\leq k\cdot 2\left(\tfrac{2et^2}{M}\right)^{M/2} \label{eq:pcb}
	\end{align}
	
	\paragraph*{Alice's Opening Attack}
	If Alice changes her mind and tries to commit to $b\oplus1$ instead, she needs to send Bob $(b\oplus1,\hat m^k)$ where $\hat m^k$ is chosen so that Bob measures the maximum amount of zeroes when doing photon counting. The state that results from displacing the original $|\psi\rangle$ by $\hat m^k$ is computed explicitly for $b=1$:
	\begin{align}
		|\tilde\psi\rangle = \otimes_{j=1}^k|\sqrt{E}(e^{\mathbbm{i}2\pi(m_j+\tfrac{1}{2})/M}-e^{\mathbbm{i}2\pi \hat m_j/M})\rangle.
	\end{align}
	We show below that Alice's optimal attack lets Bob's probability of measuring $0,\ldots,0$ equal 
	\begin{align}\label{eqn:Bob-catching-Alice}
		\mathbb P(0^k)=\exp(-E\cdot k \cdot 4\cdot \sin^2(\pi/ 2M)).
	\end{align}
	Thus Alice's cheating probability $p_{CA}$ is given by $p_{CA}=\exp(-E\cdot k \cdot 4\cdot \sin^2(\pi/ 2M))$. Upon setting e.g. $E=1$, we see that this probability decays exponentially with growing $k$. For large values of $M$ we have: $\mathbb P(0^k)\approx  1- \frac{k \pi^2}{M^2}$, ignoring higher order terms.
	
	Thus a trade-off arises: When the number of phase shifts $M$ is increased, Alice's chances of successfully implementing her attack grow. On the other hand side when $k$ is increased, Bob's chances of catching Alice when she cheats increase. 
	
	We proceed by proving \eqref{eqn:Bob-catching-Alice}: The probability that Bob measures $0$ is, per mode $j$, minimized by setting $\hat m_j=m_j$. This can be seen as follows: For two arbitrary coherent states,  $|\langle \alpha| \beta\rangle|^2 = e^{-|\alpha-\beta|^2}$ \cite[Eq. 3.33]{glauber1963coherent}. 
	Therefore, the probability $\mathbb P(0|i)$ of obtaining the measurement outcome $0$ in the $i$-th measurement is 
	\begin{align}
		\mathbb P(0|i) &= |\langle 0|\sqrt{E}(e^{\mathbbm{i}2\pi(m_j+\tfrac{1}{2})/M}-e^{\mathbbm{i}2\pi \hat m_j/M})  \rangle|^2 \\
		&= \exp(-E\cdot4\cdot \sin^2(\pi(\tfrac{m_j-\hat m_j+1/2}{M})))
	\end{align}        
	
	Let $f(m) = \sin(\pi(m+\tfrac{1}{2})/M)^2$ 
	where $m =m_j-\hat m_j$. 
	We observe that from $0\ \text{to} \ \pi/2$, the function $f(m)$ is monotone increasing. Therefore, the corresponding value which minimizes the function, $f(m)$ is for $m=0$. The value of the function becomes 
	\begin{align} \label{m= 0}
		f(m) \big|_{m=0} = \sin^{2}(\frac{\pi(m+0.5)}{M})
	\end{align}
	Also, the function is monotone decreasing from $\pi/2\ \text{to} \ \pi$. In this interval the function attains it minimum value when $m =M-1$. Its value then becomes
	\begin{align} \label{m= M-1}
		f(m) \big|_{m=M-1} 
		&= \sin^{2}(\tfrac{\pi}{2M})
	\end{align}        
	where we used $\sin{x} = \sin{(\pi-x)}$. Both \ref{m= 0} and \ref{m= M-1} are minima of function. Hence it is suffice to say, $\hat m_j=m_j$.      
	Therefore, we write the probability $\mathbb P(0|i) $ as    
	\begin{align}        
		\mathbb P(0|i)   &= \exp(-E\cdot4\cdot \sin^2(\pi/ 2M))
	\end{align}
	where we used $1- e^{i \theta} = 2 \sin{\theta/2}$.    
	
	\paragraph*{Cheating Probabilities}
	The question arises, whether the protocol can be secure both against cheating Alice and against cheating Bob. 
	Let $\epsilon>0$ be a pre-agree security parameter. The task is to find $k,M,t$ such that 
	\begin{align}
		\max\{p_{CA},p_{CB}\}\leq\epsilon.
	\end{align}
	Given the formulas \eqref{eq:pcb} and \eqref{eqn:Bob-catching-Alice}, this is equivalent to requiring
	\begin{align}
		\max\{\exp(-4t^2k\cdot \sin^2(\tfrac{\pi}{2M})), 2k\left(\tfrac{2et^2}{M}\right)^{M/2}\}\leq\epsilon.\nonumber
	\end{align}
	Assuming without loss of generality $t=1$, $\epsilon$ security is thus given provided that
	\begin{align}
		\max\{\exp(- k \cdot \sin^2(\tfrac{\pi}{2M})), k\cdot 2\left(\tfrac{2e}{M}\right)^{\tfrac{M}{2}}\}\leq\epsilon.
	\end{align}
	Since $\sin(x)\geq2x/\pi$ it is clear that $\epsilon$-security is guaranteed once both of the below inequalities hold:
	\begin{align}
		\exp(-\tfrac{ k}{M^2})&\leq\epsilon\label{eq:security-1}\\
		k\cdot 2\left(\tfrac{2e}{M}\right)^{M/2}&\leq\epsilon.\label{eq:security-2}
	\end{align}
	According to \eqref{eq:security-1} for a fixed $M\geq1$ it has to hold that $k \geq M^2\ln{\frac{1}{\epsilon}}$. According to \eqref{eq:security-2} we need $k\leq \tfrac{\epsilon}{2}(\tfrac{M}{2e})^{M/2}$. A standard choice that works for all large enough $M$ is $k=M^3$. Large values of the \emph{received} energy $t^2$ enforce larger values of $M$ and hence $k$.
	
	\subsection{Mayers Attack}\label{subsec:mayers-attack}
	\paragraph*{Description of Attack}
	For quantum \gls{bc} protocols, Mayers attack \cite{mayersImpossibility,lo1998quantum} assumes Alice will create a specific quantum state which allows her to delay her choice until the very moment where she reveals her commitment. We detail how Alice could implement such an attack in the case at hand, and give initial arguments indicati ng that implementing Mayer's attack in the current case would be extremely difficult. Executing Mayers attack requires Alice to create a purification of the states $\sigma_b$. We thus first detail the structure of these purifications:

	The states $\sigma_b$ decompose as 
	\begin{align}\label{eqn:decomposition-of-sigma-b}
		\textstyle\sigma_b=\sum_{r=0}^{M-1}\;|\phi_{r,b}\rangle\langle \phi_{r,b}|
	\end{align}
	where for $r=0,\ldots,M-1$ we have 
	\begin{align}  \label{rho_b}
		\textstyle|\phi_{r,b}\rangle
		:= e^{-\tfrac{t^2}{2}}\sum_{k=0}^{\infty}\frac{t^{\,r+kM}e^{i\pi b k}}{\sqrt{(r+kM)!}}|r+kM\rangle.
	\end{align}
	The corresponding eigenvalues are given by 
	\begin{align}
		\textstyle\lambda_r = e^{-t^2}\sum_{k=0}^{\infty}\frac{t^{\,2(r+kM)}}{(r+kM)!},
	\end{align}
	A purification of $\sigma_b$ is thus given by
	\begin{align}
		\textstyle|\Phi_b\rangle :=\sum_{r=0}^{M-1}\sqrt{\tfrac{1}{\lambda_r}}|\phi_{r,b}\rangle\otimes|\phi_{r,b}\rangle.
	\end{align}
	The purifications $|\Phi_0\rangle$ and $|\Phi_1\rangle$ are as
	\begin{align}
		|\Phi_1\rangle = \eins\otimes U|\phi_0\rangle,
	\end{align}
	where the unitary operator $U$ is given by 
	\begin{align}
		\textstyle U = \left(\sum_{r=0}^{M-1}e^{-\mathbbm{i}\tfrac{\pi\cdot r}{M}}\lambda_r^{-1}|\phi_{r,b}\rangle\langle \phi_{r,b}|\right) e^{\mathbbm{i}\frac{\pi}{M}\hat a^\dagger\hat a}.
	\end{align}

	In order to execute Mayers attack, Alice creates without loss of generality the purification $|\Phi_{0}\rangle$. Since $|\Phi_1\rangle = \eins\otimes U|\Phi_0\rangle$, she can later change her mind. 
	
	Depending on the choice she made directly before the reveal stage, she will thus hold either $|\Phi_0\rangle$ or $|\Phi_1\rangle$, and can choose a POVM accordingly. We thus describe two POVMS $\Theta_0,\Theta_1$ whose elements are defined as follows:
	\begin{align}
		|\chi_m\rangle&:=\frac{1}{\sqrt{M}}\sum_{r=0}^{M-1} e^{\,i\frac{2\pi m r}{M}}\,|\Phi_{r,0}\rangle,\\
		\theta_{0,m}&:=|\chi_m\rangle\langle\chi_m|,\qquad \theta_{0,M}:=\eins-\sum_{m=0}^{M-1}\theta_{0,m}.
	\end{align}
	Set $\theta_{1,m}:=U\theta_{0,m}U^\dagger$. Then
	\begin{align} \label{probability}
		p(m|b) &= \tr_{AB}(|\Phi_b\rangle\langle\Phi_b|\theta_{b,m})= \tfrac{1}{M}.
	\end{align}
	Further, the post measurement state is given by 
	\begin{align}
		\textstyle\sigma_{m|b}=|t e^{2\pi\mathbbm{i}\tfrac{k+b/2}{M}}\rangle\langle t e^{2\pi\mathbbm{i}\tfrac{k+b/2}{M}}|\otimes \theta_{b,m}.
	\end{align}
	Thus Bob cannot learn if Alice delayed her commitment. 
	
	\paragraph*{Physical Complexity}
	The study of complexity of continuous variable quantum optics applications differs strongly from the usual study of quantum complexity for qubits \cite{rmpBraunsteinLoock,watrousComplexity}. In the absence of an abstract computational framework, several differing perspectives are used. We will list three prominent ones:\\
	\textbf{(1) Gaussianity.} In quantum optics the first distinction line is usually between those operations which are linear transformations which are described by Hamiltonians $H=\sum_{i,j}h_{i,j}\hat a^\dagger_i\hat a_j$ with $h_{i,j}=h_{j,i}^*$ where $\hat a_i^\dagger$ is the creation operator of the $i$-th photonic mode \cite{wallsMilburn}. If a quantum state is Gaussian, then all its marginals are Gaussian as well \cite{weedbrook}. Then the $A$ and $B$ marginals of the states $|\Phi_{b}\rangle$ are both equal to $\sigma_b$, which is itself exponentially close to an ideal phase averaged coherent state. The latter is non-Gaussian \cite{phavNonGaussian}, providing a first indicator that $\sigma_b$ is non-Gaussian. The Wigner function of $\sigma_b$ is given by 
	\begin{align}
		\textstyle W_b(\beta) = \frac{1}{M}\sum_{m=0}^{M-1}e^{-|2\pi\mathbbm i\tfrac{m}{M}-\beta|^2}
	\end{align}
	and therefore not Gaussian. Thus $\sigma_b$ cannot be Gaussian, proving that the above described attack by Alice does not qualify as ``simple'' in the linear optical sense. A more detailed analysis following the methodology of \cite{nonGaussianity} is left to future work.\\
	\textbf{(2) Stellar Rank.} The stellar rank of a single mode state has been defined in \cite{stellarRepresentation}. For $|\psi\rangle=\sum_{n=0}^\infty c_n|n\rangle$ its stellar function is given by $F^*_\psi(\alpha):=e^{|\alpha|^2/2}\langle\alpha^*,\psi\rangle$. The stellar rank $r^*(\psi)$ of $\psi$ is then defined to be the the number of zeros of $F^*_\psi$, counted with multiplicity. It equals the minimum number of photon additions that are required to engineer $\ket{\psi}$ \cite{stellarRepresentation}. The stellar function of $|\phi_{r,b}\rangle$ equals 
	\begin{align}
		F^*_\psi(\alpha)
		&=e^{-\tfrac{|\alpha|^2+t}{2}}(\alpha te^{\mathbbm{i}\pi b})^{r}E_{M,r}(e^{\mathbbm i\pi b}\cdot (t\alpha)^M)
	\end{align}
	where $E_{x,y}(z)$ is the two-parameter Mittag-Leffler function $E_{x,y}(z) := \sum_{k=0}^\infty\frac{z^k}{\Gamma(x\cdot k + y)}$. The number of zeros of $E_{x,y}$ is infinite except for the cases $y=1,0,-1,-2,\ldots$ \cite{zeros}. In our case $y=r\in\mathbb N$ and $r=0,\ldots,M-1$. Therefore the stellar rank of at least one $|\Phi_{r,b}\rangle$ is infinite as soon as $M>2$. The stellar rank of a mixed state $\kappa$ is defined as $r^*(\kappa):=\inf\sum_ip_ir^*(\psi_i)$, where the infimum is taken over all decompositions $\kappa=\sum_ip_i|\psi_i\rangle\langle\psi_i|$ of $\kappa$ into pure states. Since $\sigma_b$ can be expressed both via its eigen decomposition and via its defining formula \eqref{eqn:average-code-state}, its stellar rank equals $r^*(\sigma_b)=0$. Thus from the perspective of stellar rank, the difficulty really is not to create the states $\sigma_b$ as in \eqref{eqn:average-code-state}, but in engineering Mayers attack via creation of the states $|\Phi_{r,b}\rangle$. This complexity gap illustrates clearly the simplicity of our protocol, and the hardness of breaking its security.\\    
	\textbf{(3) Specific Protocols.} In practice, we expect $\Phi_b$ to be created in an approximate sense, leaving out higher photon number contributions above a cutoff photon number $N$. In such cases, the method of creating the approximation depends on the truncated Mittag Leffler function $E_{x,y}(z) = C \sum_{k=0}^N \frac{z^k}{(x\cdot k + y-1)!}$ where $C$ is a normalization constant. For simplification, consider $x=y=1$, then $E_{x,y}(z)= 1+ z + \tfrac{z^2}{2!}+\tfrac{z^3}{3!}+ \dots +\tfrac{z^{N}}{N!}$. The highest degree of the above polynomial is $N$. Any nonzero polynomial of degree $N$ (with complex coefficients) has exactly $N$ roots in the complex plane \cite{dummit2004abstract}. Therefore the complexity of approximating $\Phi_b$ to photon number $N$ is at least $N$. For a reasonable approximation, $N\gg t^2$ is required by the properties of the Poisson distribution. The current status of the literature however is that the success probability of the most direct continuous variable quantum state creation methods decreases exponentially with $N$ \cite{creationOfQudits,clausen1999conditionalquantumstateengineering,multimodeStateEngineering}. Also indirect schemes, based on universal gate decompositions, exhibit an experimental complexity that grows fast with the polynomial order of the states or Hamiltonians to be simulated.
	\paragraph*{Impact on Implementation} In order to decrease Alice's chances of running Mayers attack, Bob should require large values of $t$. In order to fulfill $\epsilon$-security this implies $M$ and $k=M^3$ will be large. To reach reasonable values for $M$ and $k$, the transmittivity $\tau$ should therefore be guaranteed to be high by the \gls{ttp}.

	\section{Conclusion}
	We presented a novel quantum \gls{bc} protocol. We provided proof of its basic functionality in the honest but curious settings. We detailed how Alice can carry out an attack following the concept of Mayers attack. We showed that Alice, in order to be successful with this attack, needs to create high energy entangled states which cannot be created with simple linear optics. We argued why the execution of Mayers attack on our protocol is likely a hard problem, thereby providing a new means of understanding the impact of fundamental no-go results in system design. The exact complexity analysis of this attack is left to future work, since it requires in-depth analysis of possible ways of creating $|\Phi_{AB}\rangle$. We emphasize the need of developing computational models at the physical layer, which make it possible to quantify hardness of protocols such as the one proposed in this work. Our proposal is readily implementable with today's technology and provides a fresh perspective going significantly beyond state of the art assumptions that rely on the non-availability of quantum storage \cite{konig2012unconditional} or separability assumptions \cite{chaoui2025securequantumbitcommitment}.
	\section*{Acknowledgement}
	This work was financed by the DFG via grant NO 1129/2-1. The authors acknowledge the financial support by the Federal Ministry of Research, Technology and Space of Germany via grants 16KIS1598K, 16KISQ093, 16KISQ039, 16KISR026 and 16KISQ077. The generous support of the state of Bavaria via Munich Quantum Valley, the NeQuS- and the 6GQT project is greatly appreciated. The authors acknowledge the financial support by the Federal Ministry of Research, Technology and Space of Germany in the programme of “Souverän. Digital. Vernetzt.”. Joint project 6G-life, project identification number: 16KISK002.
	
	\bibliographystyle{ieeetr}
	\bibliography{bib}

@inproceedings{chailloux2011optimal,
  title={Optimal bounds for quantum bit commitment},
  author={Chailloux, Andr{\'e} and Kerenidis, Iordanis},
  booktitle={2011 IEEE 52nd Annual Symposium on Foundations of Computer Science},
  pages={354--362},
  year={2011},
  organization={IEEE}
}

@article{rabin2005exchange,
  title={How to exchange secrets with oblivious transfer},
  author={Rabin, Michael O},
  journal={Cryptology ePrint Archive},
  year={2005}
}

@article{brassard1997brief,
  title={A brief review on the impossibility of quantum bit commitment},
  author={Brassard, Gilles and Cr{\'e}peau, Claude and Mayers, Dominic and Salvail, Louis},
  journal={arXiv preprint quant-ph/9712023},
  year={1997}
}

@article{lo1998quantum,
  title={Why quantum bit commitment and ideal quantum coin tossing are impossible},
  author={Lo, Hoi-Kwong and Chau, Hoi Fung},
  journal={Physica D: Nonlinear Phenomena},
  volume={120},
  number={1-2},
  pages={177--187},
  year={1998},
  publisher={Elsevier}
}

@article{kaniewski2013secure,
  title={Secure bit commitment from relativistic constraints},
  author={Kaniewski, J{\k{e}}drzej and Tomamichel, Marco and H{\"a}nggi, Esther and Wehner, Stephanie},
  journal={IEEE Transactions on Information Theory},
  volume={59},
  number={7},
  pages={4687--4699},
  year={2013},
  publisher={IEEE}
}

@article{konig2012unconditional,
  title={Unconditional security from noisy quantum storage},
  author={Konig, Robert and Wehner, Stephanie and Wullschleger, J{\"u}rg},
  journal={IEEE Transactions on Information Theory},
  volume={58},
  number={3},
  pages={1962--1984},
  year={2012},
  publisher={IEEE}
}

@article{damgaard2008cryptography,
  title={Cryptography in the bounded-quantum-storage model},
  author={Damg{\aa}rd, Ivan B and Fehr, Serge and Salvail, Louis and Schaffner, Christian},
  journal={SIAM Journal on Computing},
  volume={37},
  number={6},
  pages={1865--1890},
  year={2008},
  publisher={SIAM}
}

@article{wehner2008cryptography,
  title={Cryptography from noisy storage},
  author={Wehner, Stephanie and Schaffner, Christian and Terhal, Barbara M},
  journal={Physical Review Letters},
  volume={100},
  number={22},
  pages={220502},
  year={2008},
  publisher={APS}
}

@article{adlam2015deterministic,
  title={Deterministic relativistic quantum bit commitment},
  author={Adlam, Emily and Kent, Adrian},
  journal={International Journal of Quantum Information},
  volume={13},
  number={05},
  pages={1550029},
  year={2015},
  publisher={World Scientific}
}

@article{glauber1963coherent,
  title={Coherent and incoherent states of the radiation field},
  author={Glauber, Roy J},
  journal={Physical Review},
  volume={131},
  number={6},
  pages={2766},
  year={1963},
  publisher={APS}
}

@article{logConvexity,
 ISSN = {00255572},
 URL = {http://www.jstor.org/stable/3620772},
 author = {David Fowler},
 journal = {The Mathematical Gazette},
 number = {501},
 pages = {433--441},
 publisher = {Mathematical Association},
 title = {The Factorial Function: Convex Functions, the Bohr-Mollerup-Artin Theorem, and Some Formulae},
 urldate = {2025-10-11},
 volume = {84},
 year = {2000}
}

@book{nielsen-chuang,
  author = {Nielsen, Michael A. and Chuang, Isaac L.},
  publisher = {Cambridge University Press},
  title = {Quantum Computation and Quantum Information},
  year = 2000
}

@Inbook{watrousComplexity,
author="Watrous, John",
title="Quantum Computational Complexity",
bookTitle="Encyclopedia of Complexity and Systems Science",
year="2009",
publisher="Springer New York",
address="New York, NY",
pages="7174--7201",
isbn="978-0-387-30440-3",
doi="10.1007/978-0-387-30440-3_428",
url="https://doi.org/10.1007/978-0-387-30440-3_428"
}

@book{wallsMilburn,
booktitle = {Quantum optics},
edition = {2nd ed.},
isbn = {3540285733},
author = {Walls, D. F. and Milburn, G. J.},
address = {Berlin},
language = {eng},
publisher = {Springer},
title = {Quantum optics / D.F. Walls, Gerard J. Milburn.},
year = {2008},
}

@article{weedbrook,
  title = {Gaussian quantum information},
  author = {Weedbrook, Christian and Pirandola, Stefano and Garc\'{\i}a-Patr\'on, Ra\'ul and Cerf, Nicolas J. and Ralph, Timothy C. and Shapiro, Jeffrey H. and Lloyd, Seth},
  journal = {Rev. Mod. Phys.},
  volume = {84},
  issue = {2},
  pages = {621--669},
  numpages = {0},
  year = {2012},
  month = {May},
  publisher = {American Physical Society},
  doi = {10.1103/RevModPhys.84.621},
  url = {https://link.aps.org/doi/10.1103/RevModPhys.84.621}
}

@article{phavNonGaussian,
author = {Alessia Allevi and Stefano Olivares and Maria Bondani},
journal = {Opt. Express},
number = {22},
pages = {24850--24855},
publisher = {Optica Publishing Group},
title = {Manipulating the non-Gaussianity of phase-randomized coherent states},
volume = {20},
month = {Oct},
year = {2012},
url = {https://opg.optica.org/oe/abstract.cfm?URI=oe-20-22-24850},
doi = {10.1364/OE.20.024850},
}

@article{approximateBC,
   title={Reexamination of quantum bit commitment: The possible and the impossible},
   volume={76},
   ISSN={1094-1622},
   url={http://dx.doi.org/10.1103/PhysRevA.76.032328},
   DOI={10.1103/physreva.76.032328},
   number={3},
   journal={Physical Review A},
   publisher={American Physical Society (APS)},
   author={D’Ariano, Giacomo Mauro and Kretschmann, Dennis and Schlingemann, Dirk and Werner, Reinhard F.},
   year={2007},
   month=sep }

@article{Chiribella_2013,
   title={A short impossibility proof of quantum bit commitment},
   volume={377},
   ISSN={0375-9601},
   url={http://dx.doi.org/10.1016/j.physleta.2013.02.045},
   DOI={10.1016/j.physleta.2013.02.045},
   number={15},
   journal={Physics Letters A},
   publisher={Elsevier BV},
   author={Chiribella, Giulio and D'Ariano, Giacomo Mauro and Perinotti, Paolo and Schlingemann, Dirk and Werner, Reinhard},
   year={2013},
   month=jun, pages={1076–1087} }

@misc{chaoui2025securequantumbitcommitment,
      title={Secure quantum bit commitment from separable operations}, 
      author={Ziad Chaoui and Anna Pappa and Matteo Rosati},
      year={2025},
      eprint={2501.07351},
      archivePrefix={arXiv},
      primaryClass={quant-ph},
      url={https://arxiv.org/abs/2501.07351}, 
}

@article{mayersImpossibility,
  title = {Unconditionally Secure Quantum Bit Commitment is Impossible},
  author = {Mayers, Dominic},
  journal = {Phys. Rev. Lett.},
  volume = {78},
  issue = {17},
  pages = {3414--3417},
  numpages = {0},
  year = {1997},
  month = {Apr},
  publisher = {American Physical Society},
  doi = {10.1103/PhysRevLett.78.3414},
  url = {https://link.aps.org/doi/10.1103/PhysRevLett.78.3414}
}

@article{rmpBraunsteinLoock,
  title = {Quantum information with continuous variables},
  author = {Braunstein, Samuel L. and van Loock, Peter},
  journal = {Rev. Mod. Phys.},
  volume = {77},
  issue = {2},
  pages = {513--577},
  numpages = {0},
  year = {2005},
  month = {Jun},
  publisher = {American Physical Society},
  doi = {10.1103/RevModPhys.77.513},
  url = {https://link.aps.org/doi/10.1103/RevModPhys.77.513}
}

@article{stellarRepresentation,
  title = {Stellar Representation of Non-Gaussian Quantum States},
  author = {Chabaud, Ulysse and Markham, Damian and Grosshans, Fr\'ed\'eric},
  journal = {Phys. Rev. Lett.},
  volume = {124},
  issue = {6},
  pages = {063605},
  numpages = {6},
  year = {2020},
  month = {Feb},
  publisher = {American Physical Society},
  doi = {10.1103/PhysRevLett.124.063605},
  url = {https://link.aps.org/doi/10.1103/PhysRevLett.124.063605}
}

@misc{zeros,
    title = {On Zeros of Functions of Mittag--Leffler Type},
    author = {Sedletskii, A.M. },
    journal = {Mathematical Notes},
    volume = {68}, 
    pages = {602--613},
    year = {2000},
    url = {https://doi.org/10.1023/A:1026671508108}
}

@article{nonGaussianity,
  title = {Quantifying the non-Gaussian character of a quantum state by quantum relative entropy},
  author = {Genoni, Marco G. and Paris, Matteo G. A. and Banaszek, Konrad},
  journal = {Phys. Rev. A},
  volume = {78},
  issue = {6},
  pages = {060303},
  numpages = {4},
  year = {2008},
  month = {Dec},
  publisher = {American Physical Society},
  doi = {10.1103/PhysRevA.78.060303},
  url = {https://link.aps.org/doi/10.1103/PhysRevA.78.060303}
}

@book{dummit2004abstract,
  title={Abstract algebra},
  author={Dummit, David Steven and Foote, Richard M and others},
  volume={3},
  year={2004},
  publisher={Wiley Hoboken}
}

@misc{clausen1999conditionalquantumstateengineering,
      title={Conditional quantum state engineering at beam splitter arrays}, 
      author={J. Clausen and M. Dakna and L. Knoell and D. -G. Welsch},
      year={1999},
      eprint={quant-ph/9905085},
      archivePrefix={arXiv},
      primaryClass={quant-ph},
      url={https://arxiv.org/abs/quant-ph/9905085}, 
}

@article{multimodeStateEngineering,
  title = {Realization of multimode operators with passive linear optics and photodetection},
  author = {Clausen, J. and Kn\"oll, L. and Welsch, D.-G.},
  journal = {Phys. Rev. A},
  volume = {68},
  issue = {4},
  pages = {043822},
  numpages = {13},
  year = {2003},
  month = {Oct},
  publisher = {American Physical Society},
  doi = {10.1103/PhysRevA.68.043822},
  url = {https://link.aps.org/doi/10.1103/PhysRevA.68.043822}
}

@article{creationOfQudits,
   title={Heralded creation of photonic qudits from parametric down-conversion using linear optics},
   volume={97},
   ISSN={2469-9934},
   url={http://dx.doi.org/10.1103/PhysRevA.97.053814},
   DOI={10.1103/physreva.97.053814},
   number={5},
   journal={Physical Review A},
   publisher={American Physical Society (APS)},
   author={Yoshikawa, Jun-ichi and Bergmann, Marcel and van Loock, Peter and Fuwa, Maria and Okada, Masanori and Takase, Kan and Toyama, Takeshi and Makino, Kenzo and Takeda, Shuntaro and Furusawa, Akira},
   year={2018},
   month=may }

\end{document}